\documentclass[12pt]{article}
\usepackage{amsfonts}
\usepackage{cite}
\usepackage{bbm}
\usepackage{amsmath,amssymb}
\usepackage{graphicx,color}
\usepackage{subfigure}

\renewcommand{\thesection}
{\arabic{section} \hspace{-.5em}
}
\renewcommand{\thesubsection}
{\arabic{section}.\arabic{subsection}   \hspace{-.5em}
}
\renewcommand{\thesubsubsection}
{\arabic{section}.\arabic{subsection}.\arabic{subsubsection} \hspace {-.5em}
 }

\makeatletter
\renewcommand\section{
\@startsection{section}{3}{\z@}%
{-4.5ex\@plus -1ex \@minus -.2ex}%
{1.5ex \@plus .2ex}%
{\normalfont\large\bfseries\mathversion{bold}}}
\renewcommand\subsection{
\@startsection{subsection}{3}{\z@}%
{-3ex\@plus -1ex \@minus -.2ex}%
{0.7ex \@plus .2ex}%
{\normalfont\normalsize\bfseries\mathversion{bold}}}
\renewcommand\subsubsection{
\@startsection{subsubsection}{3}{\z@}%
{-3.25ex\@plus -1ex \@minus -.2ex}%
{1.5ex \@plus .2ex}%
{\normalfont\normalsize\itshape}}
\makeatother

\makeatletter \@addtoreset{equation}{section} \makeatother
\renewcommand{\theequation}{\arabic{section}.\arabic{equation}}

\makeatletter
\renewcommand{\appendix}{
\renewcommand{\thesection}{\Alph{section}  \hspace{-.5em}}
\renewcommand{\thesubsection}
{\Alph{section}.\arabic{subsection} \hspace{-.5em}}
\renewcommand{\thesubsubsection}
{\Alph{section}.\arabic{subsection}.\arabic{subsubsection} \hspace  {-.5em}}
\@addtoreset{equation}{subsection}
\renewcommand{\theequation}{\Alph{section}.\arabic{equation}}
\setcounter{section}{0}}
\makeatother

\renewcommand{\thefootnote}{\fnsymbol{footnote}}

\def\bc       {\begin{center}}
\def\ec       {\end{center}}

\def\ket  {\rangle}

\def\({\left(}
\def\){\right)}
\def\bs#1{\boldsymbol{#1}}
\newcommand{\hlambda}{{\hat{\lambda}}}
\DeclareMathOperator{\im}{Im}

\newcommand{\nn}{\nonumber}

\newcommand{\eqb}{\begin{eqnarray}}
\newcommand{\eqe}{\end{eqnarray}}
\def\comma      { \, , }
\def\period     { \, . }

\def\calO   {{\cal O}}
\newcommand{\bbR}{{\mathbb R}}
\newcommand{\bbZ}{{\mathbb Z}}

\def\tilM   {\tilde{M}}

\begin{document}
\def\papertitlepage{\baselineskip 3.5ex \thispagestyle{empty}}
\def\preprinumber#1#2#3{\hfill \begin{minipage}{2.6cm} #1
                \par\noindent #2
              \par\noindent #3
             \end{minipage}}
\renewcommand{\thefootnote}{\fnsymbol{footnote}}
\newcounter{aff}
\renewcommand{\theaff}{\fnsymbol{aff}}
\newcommand{\affiliation}[1]{
\setcounter{aff}{#1} $\rule{0em}{1.2ex}^\theaff\hspace{-.4em}$}
%
%
\papertitlepage
\setcounter{page}{0}
\preprinumber{}{UTHEP-654}{}
\baselineskip 0.8cm
\vspace*{2.5cm}
\begin{center}
{\Large\bf Gluon scattering amplitudes   \vspace*{0.5ex} \\
from gauge/string duality and integrability}
\end{center}
\vskip 4ex
\baselineskip 0.7cm
\begin{center}
       
         Yuji  ~Satoh\footnote[3]{\tt ysatoh@het.ph.tsukuba.ac.jp}

\vskip 2ex
 
    {\it Institute of Physics, University of Tsukuba} \\
    {\it Tsukuba, Ibaraki 305-8571, Japan}
\end{center}
\vskip 11ex
%
\baselineskip=3.5ex

\begin{center} {\bf Abstract} \end{center}

\par\medskip
\ 
We discuss gluon scattering amplitudes/null-polygonal Wilson loops
 of ${\cal N} = 4$ super Yang-Mills theory at strong coupling based 
 on the gauge/string duality and its underlying
 integrability. We focus on the amplitudes/Wilson loops corresponding 
 to the minimal surfaces in $AdS_{3}$, which
 are described by the thermodynamic Bethe ansatz
 equations of the homogeneous sine-Gordon model.
 Using conformal perturbation theory and an interesting relation 
 between the $g$-function (boundary entropy) and the T-function, 
 we derive analytic expansions around the limit where the Wilson loops
 become regular-polygonal. 
 We also compare our analytic results with those at two loops, to find that
 the rescaled remainder functions are close to each other 
 for all multi-point amplitudes.

%
%
%
%
%

\vspace*{\fill}
\noindent
July 2012%
\footnote[0]{
Contribution to the proceedings of ``Progress in Quantum Field Theory and String Theory'', 
April 3-7, 2012, Osaka City University, Osaka, Japan.}

\newpage
\renewcommand{\thefootnote}{\arabic{footnote}}
\setcounter{footnote}{0}
\setcounter{section}{0}
\baselineskip = 3.3ex
\pagestyle{plain}

\section{Introduction}

In this talk, we would like to discuss gluon scattering amplitudes of four-dimensional 
${\cal N} =4$ super Yang-Mills  theory (SYM) at strong coupling 
by using the gauge/string duality and its integrability. This talk is based 
on Refs. \cite{Hatsuda:2010cc,Hatsuda:2010vr,Hatsuda:2011ke,Hatsuda:2011jn}.

Let us start this talk by a rather general introduction to this subject. 
Already almost ten years ago,
some integrability was discovered in the gauge/string duality or 
the AdS/CFT correspondence \cite{Minahan:2002ve,Bena:2003wd}.
This discovery of the integrability opened up new dimensions in the study 
of the gauge/string duality. For example,  owing to the integrability, one can now 
discuss the gauge/string  duality beyond supersymmetric sectors in a precise 
manner. Moreover, one can now quantitatively analyze strong-coupling 
dynamics of a gauge theory by using the integrability. Besides,  
by deeply understanding the gauge/string duality,
one may expect to obtain useful insights into a lot of applications. 

In fact, very impressive results 
 about the spectrum of the planar AdS/CFT 
correspondence have been obtained \cite{Beisert:2010jr}.%
\footnote{See also the contributions to the proceedings 
by G.E.  Arutyunov, V. Kazakov and R. Suzuki.
}
Namely, one can now calculate/analyze the spectrum of single-trace operators 
of ${\cal N} = 4$ SYM 
or of the superstrings on $AdS_{5} \times S^{5}$ for an arbitrary 't Hooft coupling 
in the planar limit. The spectrum is obtained by solving a set of integral equations 
of the type called the thermodynamic Bethe ansatz (TBA) 
equations \cite{Zamolodchikov:1989cf},  which typically appear
in the study of finite size effects of two-dimensional integrable models.
The spectrum has also been checked  for simple operators  
on the weak-coupling side
up to five loops \cite{Bajnok:2009vm,Eden:2012fe}. The results are  really remarkable.

Given this success on the spectrum, one may well expect that the integrability 
should be useful also for other aspects or applications of the gauge/string duality. 
It has turned out that this is indeed the case, and we can now analyze gluon 
scattering amplitudes/null-polygonal Wilson loops of ${\cal N} =4$ SYM at strong 
coupling by using the integrability.
An overview is as follows: By the AdS/CFT correspondence, the scalar part of the 
the maximally-helicity-violating (MHV) gluon scattering amplitude%
\footnote{
Regarding the MHV amplitude, see also the contribution by T.R. Taylor.
}
 at strong coupling 
is represented 
 by the area of the minimal surfaces in $AdS_{5}$ which have 
a null-polygonal boundary on the boundary of $AdS_{5}$ \cite{Alday:2007hr}. 
These minimal surfaces also give the expectation values of the Wilson loops 
along the null-polygonal boundary according to the AdS/CFT correspondence,
implying an equivalence between the amplitudes and the Wilson loops. 
The minimal surfaces are then described 
by a set of integral equations of the TBA type due to 
the integrability \cite{Alday:2009yn,Alday:2009dv,Alday:2010vh,Hatsuda:2010cc}.
Schematically,

\vspace{0.5ex}
\begin{center}
\fbox{\parbox{8.7cm}{ \hspace{8.5ex}  Gluon scattering amplitudes/ \\ 
\hspace*{0.9ex} null-polygonal Wilson loops  at strong coupling \ \ }}

\vspace{1.5ex}  \hspace{17ex}
$\Uparrow$  \qquad   {\it AdS/CFT} 

\vspace{1.5ex}
\fbox{\parbox{5.2cm}{ \hspace{1ex} Minimal surfaces in $AdS_5 $  \ }}

\vspace{1.5ex} \hspace{18.7ex}
 $\Uparrow$  \qquad {\it Integrability}  
 
\vspace{1.5ex}
{\fbox{\parbox{7.8cm}{  \hspace{1ex} Thermodynamic Bethe ansatz equations \ }} }

\end{center}
\vspace{0.5ex}

\noindent
The TBA equations for the amplitudes/Wilson loops 
are different from those for the spectral problem.
However, it is interesting to see that TBA equations appear again in a different 
context of the AdS/CFT correspondence. 

In the following,  we discuss the MHV amplitudes/null-polygonal 
Wilson loops of ${\cal N} =4$ SYM
at strong coupling by using the underlying two-dimensional integrable models and 
conformal filed theories (CFTs). In particular, we derive analytic expansions 
of the amplitudes/Wilson loops
around certain kinematic points corresponding to regular-polygonal Wilson loops.
We focus on the case where the  momenta of the external particles are contained in a 
two-dimensional subspace of four-dimensional Minkowski space, in other words, 
where the minimal surfaces are included in $AdS_{3} $ within $AdS_{5}$.
We also compare our analytic results with those at two loops, and find that 
the rescaled remainder functions are close to each other 
for all multi-point amplitudes.

Before moving on to the discussion below, 
we would like to mention that the integrability has recently been
applied also to the  computation of the correlation functions and the quark anti-quark 
potentials (or the cusp anomalous dimensions).%
\footnote{
See also the contribution by S. Komatsu.}
 The application of the integrability is now
being expanded in this way. We would also like to mention that the development 
on the strong-coupling/string 
side is stimulating the study of ${\cal N} =4$ SYM at weak coupling.

The plan of the rest of this talk is as follows. In section 2, we briefly summarize
the gluons scattering amplitudes/Wilson loops of ${\cal N} =4$ SYM at strong coupling 
in the context of the AdS/CFT correspondence. In section 3, we summarize how 
the amplitudes at strong coupling are obtained by using the integrability 
or the TBA equations. In section 4, which is the main part of this talk, we derive 
the analytic 
expansions of the amplitudes at strong coupling by using the two-dimensional 
integrable models and CFTs associated with the minimal surfaces.  In section 5,
we summarize the results of the lower-point cases of the eight- and ten-point amplitudes.
In section 6, we compare our analytic results at strong coupling with those at two loops.  
We conclude with a summary and discussion in section 7.

\section{Gluon scattering amplitudes at strong coupling}

We consider gluon scattering amplitudes of the four-dimensional  
maximally supersymmetric, i.e., ${\cal N} =4$,  
$SU(N_{c})$ SYM in the planar limit $N_{c} \to \infty$ 
with the 't Hooft coupling $ \lambda :=g_{YM}^{2} N_{c}$ being fixed. 
For a review, see for example Ref. \cite{Alday:2008yw}. 
Since ${\cal N} =4$ SYM is a massless gauge theory,    
the amplitudes contain infrared divergences.
These  divergences are first to be regularized, but cancel each other
in  infrared-safe quantities.

In this talk, we are interested in a  class of the amplitudes where all the helicity
is the same except for two particles, i.e., the MHV amplitude. 
In the planar limit, one can factorize the tree amplitude
from the MHV amplitude. 
By the AdS/CFT correspondence, the remaining scalar part of the amplitude 
at strong coupling is given by the (regularized) area, $A$,  of
minimal surfaces in $AdS_{5}$ \cite{Alday:2007hr}:
\eqb\label{MArea}
    {\cal M} \sim e^{-\frac{\sqrt{\lambda}}{2\pi} A } \period
\eqe
In Fig.~1, we show an example of a minimal surface representing a four-point amplitude.
\begin{figure}[t]
  \bc
   \includegraphics*[width=4.3cm]{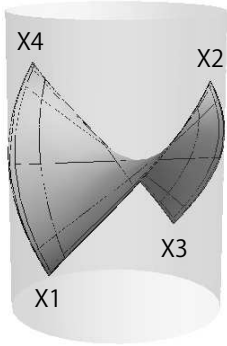}
  \vspace*{2pt}
 \caption{Null-polygonal minimal surface for four-point amplitude. 
The $AdS$ space is represented by the solid cylinder, whereas
the $AdS$ boundary by the side of the cylinder.
}
\ec
\end{figure}
As shown in the figure, the surface extends to the $AdS$ boundary, where the surface 
has a null-polygonal boundary and cusps. The sides of the surface, 
or the differences of the cusp 
points, correspond to the momenta of the external particles,
\eqb\label{cusp-momenta}
   x_{i+1}^{\mu} - x_{i}^{\mu} = 2\pi k_{i}^{\mu} \period 
\eqe

Thus, roughly speaking, the $\tilde{n}$-point amplitudes at strong coupling are
the $\tilde{n}$-cusp minimal surfaces in $AdS$.
Since the minimal surfaces
give expectation values of the Wilson loops at strong coupling along the boundary
via the AdS/CFT correspondence, the above formula implies an equivalence 
of the MHV amplitude and the null-polygonal Wilson loops. This has  been 
confirmed also at weak coupling \cite{Drummond:2007aua}.
Consequently, we may say that we are considering
the Wilson loops of ${\cal N} =4$ SYM instead of the amplitudes.

In Ref. \cite{Alday:2007hr}, four-point amplitudes were
computed according to (\ref{MArea}). A precise agreement was then found
with the structure predicted by the Bern-Dixon-Smirnov (BDS) 
conjecture \cite{Bern:2005iz},
which proposes the form of the MHV amplitude to all orders in perturbation.
It turned out, however,  that the BDS formula
is modified  for higher-point amplitudes \cite{Alday:2007he}. 
This deviation from the BDS formula
is now called the remainder function. Because of the anomalous dual conformal 
Ward identities \cite{Drummond:2007au},
this remainder function should be a function of the cross-ratios 
of the cusp coordinates in (\ref{cusp-momenta}), which correspond to 
 the momenta of the particles. The dual conformal symmetry also assures that
 the BDS formula is exact for $ \tilde{n} \leq 5$. At $\tilde{n}=6$, the existence 
 of the remainder function has indeed been 
 confirmed \cite{Bern:2008ap,Drummond:2008aq}.
 
 Once given the BDS formula, to compute the amplitudes is the same as to compute
 the remainder function. Thus, our task is to determine the remainder function 
 as a function of the cross-ratios of $x_{i}^{\mu}$.

\section{Scattering amplitudes from TBA system}
                                                                                
After the work of Ref. \cite{Alday:2007hr}, 
there were attempts at extending  the strong-coupling computation 
to higher-point amplitudes. 
Though a special six-cusp
minimal surface has been constructed \cite{Sakai:2009ut,Sakai:2010eh},
it turned out that it is very difficult to construct
the minimal surfaces with a null-polygonal boundary.
Remarkably, it was however shown in Ref. \cite{Alday:2009yn} that
one can calculate the area of the minimal surfaces by using integrability 
without knowing the explicit form of the surfaces.

In the following, we would like to explain how to carry out this program,
focusing on the case where the minimal surfaces are contained in 
$AdS_{3} \subset AdS_{5}$ for simplicity. This corresponds to the case where
the momenta of the external particles are included in a two-dimensional 
subspace $\bbR^{1,1}$
of four-dimensional Minkowski space. Thus, we are still considering four-dimensional
physics but with special kinematics. In this case, to satisfy the momentum conservation,
the number of the external particles or the cusp points becomes even: $\tilde{n} = 2n$.
Here, we also introduce two light-cone coordinates
$x^{\pm}$ on the boundary of $AdS_{3}$.
The boundary of the minimal surfaces is then parameterized as in Fig.~2.
The boundary of the surface closes at infinity in these coordinates.
We note that  the amplitudes depend on the momenta through
the cross-ratios of $x^{\pm}_{i}$, though.
\begin{figure}[t]
 \bc
   \includegraphics*[width=4.7cm]{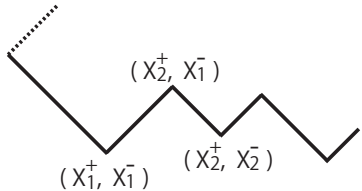}
 \vspace*{5pt}
 \caption{Boundary of null-polygonal minimal surface in $AdS_{3}$.}
 \ec
\end{figure}
\subsection{TBA equations for  minimal surfaces in $AdS_{3}$}

In Ref. \cite{Alday:2009yn}, the amplitudes corresponding to  
the minimal surfaces in $AdS_{3}$ were analyzed. 
The classical string equations of motion describing the minimal surfaces were reduced 
to the $su(2)$ Hitchin system or a generalized sinh-Gordon equation via 
the Pohlmeyer reduction. The regularized area was divided into several terms,
and they were explicitly evaluated for the eight-point amplitudes with the help of 
the results  on  the moduli space of a four-dimensional  
${\cal N} = 2$ supersymmetric theory \cite{Gaiotto:2008cd}. 
The explicit form of the surfaces is not needed in this analysis, as mentioned.

The analysis in Ref. \cite{Alday:2009yn} was extended to the minimal surfaces 
in $AdS_{5}$ in Ref. \cite{Alday:2009dv},
where the Pohlmeyer reduction led to the $su(4)$ Hitchin system. Interestingly,
it was shown there that a non-trivial part of the regularized area for the six-point
amplitudes is obtained by solving a set of integral equations, which turn out 
to be the TBA equations associated with the $\bbZ_{4}$-symmetric two-dimensional 
integrable model \cite{Koberle:1979sg}.  Following this work, the integral 
equations for the 10- and 12-cusp minimal surfaces in $AdS_{3}$ were derived in 
Ref. \cite{Hatsuda:2010cc}.
The general case of the $\tilde{n}$-cusp minimal surfaces in $AdS_{5}$ was 
studied in Ref. \cite{Alday:2010vh}. 
For the minimal surfaces in $AdS_{3}$ and $AdS_{4}$, 
the integral equations obtained there were identified in Ref. \cite{Hatsuda:2010cc} 
with the TBA equations of a two-dimensional integrable model
called the homogeneous-sine Gordon (HSG) model \cite{FernandezPousa:1996hi}.
In the rest of this section, we follow the results in Ref. \cite{Alday:2010vh}.

In order to calculate the $2n$-point amplitudes corresponding to the minimal surfaces
in $AdS_{3}$, one first needs to solve the following set of integral equations:
\eqb\label{AdS3TBA}
  \log \tilde{Y}_s(\theta) = - |m_s|  \cosh \theta 
   + \sum_{r=1}^{n-3} K_{sr} \ast \log(1+\tilde{Y}_r) \comma
\eqe 
where $s=1, ..., n-3$, $\tilde{Y}_{0} = \tilde{Y}_{n-2} =0$, and  $\ast$ stands 
for the convolution, $f \ast g := \int  f(\theta-\theta') g(\theta') \, d\theta' $.
The parameter $\theta$ is an auxiliary parameter called the spectral parameter.
The  complex parameters $m_{s} = |m_{s}| e^{i\varphi_{s}}$  
correspond to the shape of the minimal surfaces, which
gives the momenta of the external particles through (\ref{cusp-momenta}). 
By tildes, we denote the shift of the argument of the Y-functions, $Y_{s}(\theta)$, as
$\tilde{Y}_{s}(\theta) := Y_{s}(\theta + i \varphi_{s})$. 
The kernels in the convolution are 
\eqb
   K_{sr}(\theta) = \frac{I_{sr}}{2\pi \cosh(\theta + i\varphi_{s} - i \varphi_{r}) } 
   \comma  
\eqe
where $I_{sr} = \delta_{s, r+1} + \delta_{s, r-1}$ is  the incidence matrix 
of the $A_{n-3}$ algebra.
Thus, the ``interaction'' among the Y-functions are  
represented by the Dynkin
diagram of $A_{n-3}$ (see Fig.~3).
\begin{figure}[t]
   \bc
   \includegraphics*[width=4cm]{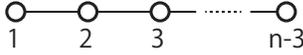}
 \vspace*{5pt}
 \caption{Dynkin diagram representing interaction among $Y_{s}$.}
 \ec
\end{figure}
Precisely, the form of  (\ref{AdS3TBA}) 
is valid for $| \varphi_{s} - \varphi_{s \pm 1}| < \pi/2$.
Outside this range, it is obtained by continuation,  
to pick up contributions of the poles in the kernels.
Under the condition that  $\tilde{Y}_{s}(\theta)$ are analytic for 
   $ | \im \theta \, | < \pi/2$, the  integral equations
are converted into a set of algebraic equations called
the Y-system \cite{Zamolodchikov:1991et}:
\eqb\label{Ysystem}
    Y_{s}^{[+1]}(\theta) Y_{s}^{[-1]}(\theta) 
    = \Bigl( 1+Y_{s+1}(\theta) \Bigr) \Bigl( 1+Y_{s-1}(\theta) \Bigr) \comma
\eqe
where the bracket  stands for the shift of the argument,
\eqb
    f^{[k]}(\theta) := f\Bigl( \theta + \frac{\pi i}{2}k  \Bigr) \period 
\eqe
The Y-functions   
represent the cross-ratios of the cusp coordinates extended 
by the spectral parameter.  Indeed, one finds that
\eqb\label{Ycrossratio}
   Y_{2r+1}^{[-1]}(0) &=& \frac{x^{+}_{r,-r-1} x^{+}_{r+1,-r-2}}{x^{+}_{r,r+1}x^{+}_{-r-2,-r-1}}
   \comma \quad 
   Y_{2r+1}^{[0]}(0) = \frac{x^{-}_{r,-r-1} x^{-}_{r+1,-r-2}}{x^{-}_{r,r+1}x^{-}_{-r-2,-r-1}}
   \comma \nn \\
   Y_{2r}^{[0]}(0) &=& \frac{x^{+}_{r,-r} x^{+}_{r+1,-r-1}}{x^{+}_{r,r+1}x^{+}_{-r-1,-r}}
   \comma \quad \hspace{4.1ex}
   Y_{2r}^{[1]}(0) = \frac{x^{-}_{r,-r} x^{-}_{r+1,-r-1}}{x^{-}_{r,r+1}x^{-}_{-r-1,-r}} \comma
\eqe
where $x_{i,j}^{\mu} := x_{i}^{\mu} - x_{j}^{\mu}$, and the indices for the cusps 
are labeled modulo $n$. 
For example, $ Y_1(-\pi i/2) = x^+_{15} x^+_{67}/x^+_{56} x^+_{17}$,
$Y_1(0) = {x^-_{15} x^-_{67}}/{x^-_{56} x^-_{17}} $ for $n=7$.
Graphically, these cross-ratios  are represented by the tetragons inside the $n$-gons 
formed by $x_{i}^{\pm}$. 
In Fig.~4, we show an example of $n=7$.
\begin{figure}[t]
 \bc
  \includegraphics*[width=5.6cm]{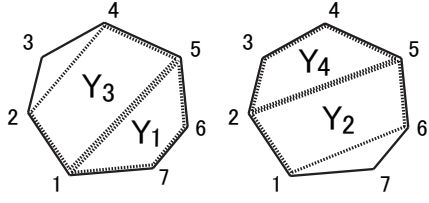}
 \vspace*{5pt}
 \caption{Graphical representation of Y-functions for $n=7$. 
 The $i$-th vertex stands for $x_{i}^{+}$ or $x_{i}^{-}$. 
  At special values  of $\theta$, $Y_{s}$ give
  the cross-ratios formed by the cusp coordinates at the corners
 of the corresponding tetragons. We omit the shifts $[k]$ of  $Y^{[k]}_{s}$.}
 \ec
\end{figure}

As mentioned above, the equations (\ref{AdS3TBA}) are nothing but the TBA equations 
of the HSG model. In the context of the two-dimensional integrable model, $m_{s}$ are
the (complexified) mass parameters, $\theta$ 
is the rapidity, and $Y_{s}$ are the exponentials of the pseudo-energies.

\subsection{Remainder function}
Once the Y-functions are obtained by solving the TBA equations (\ref{AdS3TBA}),
one can write down the formula of  the remainder function, which is defined
by the difference between the amplitude and the BDS formula. 
For the time being, we focus on the case of 
the $2n$-point amplitudes with $n$ odd. In this case, the formula reads
\eqb\label{R2n}
R_{2n} &:= & A\mbox{\small\rm (amplitude)} 
 -  A_{\rm BDS}\mbox{\small\rm (BDS formula)} \nn \\
  &= &  {7\pi\over12}(n-2) +A_{\rm periods}+\Delta A_{\rm BDS} +A_{\rm free}  \period
\eqe
The overall coupling constant $\sqrt{\lambda}$ has been omitted above.
The first term is a constant. The second term, which  is given by
\eqb\label{Aperiods}
  A_{\rm periods} = -\frac{1}{4} m_r \, I^{-1}_{rs}\,  \overline{m}_s \comma
\eqe
comes from  period integrals over an auxiliary 
hyperelliptic  curve $y^{2} = p(z)$ with  $p(z)$ being
a polynomial of degree $n-2$. 
The third term, which  is given by
\eqb
   \Delta A_{\rm BDS} = \frac{1}{4} \sum_{i,j = 1}^{n} \log\frac{ c_{i,j}^{+} }{ c_{i,j+1}^{+} } 
\log \frac{ c_{i-1,j}^{-} }{c_{i,j}^{-} }  \comma
\eqe
comes from the difference between a term 
satisfying the anomalous dual conformal Ward identities
and a similar term in the BDS formula.
Here,  $c_{i,j}^{\pm}$ are the sequential cross-ratios formed by the 
nearest neighbor distances of $x_{i}^{\pm}$,
\eqb
   c^{\pm}_{i,j} := \frac{ x^{\pm}_{i+2,i+1} x^{\pm}_{i+4,i+3} \cdots x^{\pm}_{j,i} }{
 x^{\pm}_{i+1,i} x^{\pm}_{i+3,i+2} \cdots x^{\pm}_{j,j-1} }  \period 
\eqe
Similarly to $Y_{s}$ in Fig.~4, these are graphically represented by polygons 
 inside the $n$-gons formed by $x_{i}^{\pm}$. 
In Fig.~5, we show an example of $c^{\pm}_{1,6}$ for $n=7$.
\begin{figure}[t]
   \bc
    \includegraphics*[width=2.6cm]{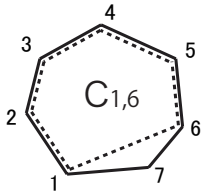}
  \vspace*{5pt}
 \caption{Graphical representation of  cross-ratios $c_{1,6}^{\pm}$ for $n=7$.
     We omit superscripts $\pm$.}
   \ec  
\end{figure}
The last term is given by the following integral,
\eqb
   A_{\rm  free}=\sum_{s=1}^{n-3} \int_{-\infty}^\infty \frac{d\theta}{2\pi} 
  | m_s | \cosh \theta \log\bigl(1+\tilde{Y}_s(\theta)\bigr) \period 
\eqe
Interestingly, this is nothing but the free energy associated with the TBA system 
(up to a factor). From these expressions, we find that 
there are two non-trivial parts, $\Delta A_{\rm BDS}$ and $A_{\rm free}$, 
in expressing the remainder function in terms of the mass parameters $m_{s}$.

Given the formula of the remainder function, what we should do is to solve 
the TBA equations.  They are solved numerically by iterations,  
to get numbers of $R_{2n}$ for given $m_{s}$.  
The TBA equations
(\ref{AdS3TBA}) are also solved exactly in the limit where all $m_{s} $  approach  0 
or $\infty$. The former
limit corresponds to the UV limit where the two-dimensional integrable model 
reduces to a CFT, or to the limit where the corresponding Wilson loops become
regular-polygonal. The latter corresponds to the IR limit where the particles 
in the two-dimensional integrable model become free massive particles, 
or to soft limits of the SYM theory.

However, the solutions to the TBA system (\ref{AdS3TBA}) have not been 
fully understood yet. 
To find the momentum dependence of the remainder function and its structures,
one also needs to find the relation between $m_{s}$ and the momenta, 
since the TBA equations are solved for given $m_{s}$.
Here, one may need some analytic data.  In addition, 
interesting analytic results  have been 
obtained at two loops 
\cite{DelDuca:2010zg,Goncharov:2010jf,DelDuca:2010zp,Heslop:2010kq}.
Taking these into account,  it would be worthwhile to explore analytic results 
at strong coupling besides in the special limits $m_{s} \to 0, \infty$.

This is the subject of the rest of this talk. In particular, we discuss analytic
expansions of the remainder function around the UV/CFT limit $m_{s} \to 0$.
For later use,  here we introduce 
the mass scale $M$, the length scale $L$ which corresponds to the inverse temperature
of the TBA system, the dimensionful mass parameters $M_{s}$ 
and the relative masses $\tilde{M}_{s} $, so that 
\eqb
  m_{s} = M_{s} L = \tilde{M}_{s} ML \period 
\eqe  
In terms of these parameters, the CFT limit we consider is represented  as
\eqb
    l := ML \to 0 \period 
\eqe

\section{Analytic expansions of amplitudes}

Now, let us move on to the discussion on the analytic 
expansion \cite{Hatsuda:2011ke,Hatsuda:2011jn}.
A basis of the expansion is the fact 
that the integral equations for 
the minimal surfaces in $AdS_{3}$ are the TBA equations of the HSG model
associated with the coset $ \widehat{su}(n-2)_{2}/ [\widehat{u}(1)]^{n-3}$ 
\cite{Hatsuda:2010cc} . 
Schematically,

\vspace{1ex}
\begin{center}    
  \fbox{\parbox{4.3cm}{ \hspace{1.6em} \ TBA for $AdS_{3}$
  \\  \hspace*{1.1em}
    $2n$-pt. amplitudes }} 
    \qquad $\Longleftarrow$ \qquad 
                  \hspace*{-2ex}  \fbox{\parbox{4.8cm}{ \hspace{0.9em} HSG from  
                  $ \displaystyle { \widehat{su}(n-2)_{2} \over  [\widehat{u}(1)]^{n-3} }$
          } }        
\end{center}       
\vspace{1ex} 
Precisely, in the TBA equations for the minimal surfaces, the resonance parameters
are purely imaginary. Though  physical interpretation of the imaginary resonance 
parameters is not clear at present, they are irrelevant to the following discussion: 
they are first set to be zero and, after the expansion is obtained, they are recovered 
so as to  maintain symmetries.  
Vanishing imaginary resonance parameters correspond to real mass parameters.
We thus consider real $m_{s}$ for the time being.

Similarly, the integral equations for the $\tilde{n}$-cusp  minimal surfaces in $AdS_{4}$
are  the TBA equations of the HSG model associated 
with the coset $\widehat{su}(\tilde{n}-4)_{4}/[\widehat{u}(1)]^{\tilde{n}-5}$ 
\cite{Hatsuda:2010cc}.
For the $AdS_{5}$ case,
the corresponding integrable model for the six-cusp minimal surfaces is
the $\bbZ_{4}$-integrable model with a twist \cite{Alday:2009dv,Hatsuda:2010vr}, 
which is equivalent to the HSG model associated with 
$\widehat{su}(2)_{4}/[\widehat{u}(1)] $ or 
 corresponding  complex sine-Gordon model with a twist. However,
the integrable model for the general $AdS_{5}$ case has not  been identified. 

The HSG model associated with the coset $ \widehat{su}(n-2)_{2}/ [\widehat{u}(1)]^{n-3} $ 
for the $AdS_{3}$ minimal surfaces is obtained by 
an integrable deformation of the corresponding coset CFT/gauged WZNW model. 
The action is  then given by
\eqb\label{SHSG}
   S_{\rm HSG} = S_{\rm gWZNW} + \beta \int d^{2}x \, \Phi \comma
\eqe
where the first term on the right-hand side is the action of the gauged WZNW model 
and $\Phi(\tilde{M}_{s})$ is a linear combination of the weight 0 adjoint operators, 
which has the coefficients depending on $\tilde{M}_{s}$ and conformal weights 
\eqb
 \Delta = \bar{\Delta} = (n-2)/n \period 
\eqe 
On dimensional grounds, the coupling $\beta$ takes the form  
\eqb
\beta = -\kappa_{n} M^{2(1-\Delta)}\comma 
\eqe
where $\kappa_{n}$ is  the dimensionless coupling.

Once we find the relation among the integral equations for the minimal surfaces, 
the two-dimensional integrable model and the corresponding CFT in the CFT/UV limit, 
we can expand the remainder function around the CFT limit by  
conformal perturbation theory (CPT). Here, we recall that there are two non-trivial parts 
in expanding $R_{2n}$, i.e.,  $A_{\rm free}$ and $\Delta A_{\rm BDS}$.
 Let us discuss their expansions  one by one.

\subsection{Expansion of free energy part}

Since  $A_{\rm free}$ is the free energy  of the HSG model,
it is straightforward to expand it around the CFT limit by using the action (\ref{SHSG})
and CPT. By the standard procedure \cite{Zamolodchikov:1989cf}, one has
\eqb
   A_{\rm free} = \frac{\pi}{6} c_{n} + f_{n}^{\rm bulk} 
   + \sum_{p=2}^{\infty} f_{n}^{(p)} l^{4p/n} \period 
\eqe
The first term $c_{n} = (n-2)(n-3)/n$ is the central charge of the coset CFT for 
$ \widehat{su}(n-2)_{2}/ [\widehat{u}(1)]^{n-3} $. The second term
 $f_{n}^{\rm bulk}$ is the bulk term given for $n$ odd by 
\eqb
   f_{n}^{\rm bulk} = \frac{1}{4} m_{r} I^{-1}_{rs} \overline{m}_{s} \comma 
\eqe
which cancels with $A_{\rm periods}$ in (\ref{Aperiods}). The terms in the summation
come from the perturbation by $\Phi$, 
\eqb
   f_{n}^{(p)} =  {\kappa_{n}^{p} \over p!} (2\pi)^{2  + 2(\Delta-1)p} \int
    \Big\langle \Phi(x_{1}) \cdots \Phi(x_{p})  \Big\rangle_{\rm c} 
    \prod_{i=2}^{p} |x_{i}|^{-2(\Delta -1)}
    d^{2}x_{i} \comma 
\eqe
where the correlators are connected ones evaluated at the CFT point and  
we have set $x_{1} = 1$. At the lowest order, we have
\eqb
   f^{(2)}_{n} = \frac{\pi}{6} C_{n}^{(2)} \kappa_{n}^{2} G^{2}(\tilde{M}_{s}) \comma
\eqe
where
\eqb
  C_{n}^{(2)} = 3 (2\pi)^{\frac{2(n-4)}{n}} 
  \gamma^{2}\Bigl(\frac{n-2}{n}\Bigr)\gamma\Big(\frac{4-n}{n}\Big) \comma
\eqe
 $\gamma(x) := \Gamma(x)/\Gamma(1-x)$, and $G(\tilde{M}_{s})$ is a normalization 
 factor defined through 
 \eqb
  \Big\langle \Phi(x) \Phi(0)\Big\rangle = { G^{2}(\tilde{M}_{s}) \over |x|^{4\Delta} } 
  \period
\eqe  
 To determine the dependence on $\tilde{M}_{s}$, one needs to find the precise 
 form of $\Phi (\tilde{M}_{s})$ in terms of the weight 0 adjoint operators. 
 We will come back to this issue later. 

\subsection{$\Delta A_{\rm BDS}$ and T-functions}

Next, let us consider the expansion of $\Delta A_{\rm BDS}$, which is given 
by the sequential cross-ratios $c_{i,j}^{\pm}$. Interestingly,  
with the help of graphical representations as in Figs. 4 and 5, 
one can show that these cross-ratios are
directly expressed by the T-functions \cite{Kuniba:2010ir},  
which are related to the Y-functions as 
\eqb\label{YT}
     Y_{s}(\theta) = T_{s+1}(\theta) T_{s-1}(\theta) \comma
     \quad 1+Y_{s}(\theta) = T^{[+1]}_{s}(\theta) T^{[-1]}_{s} (\theta) \period
\eqe
One then finds for $n$ odd  that 
\eqb
   c_{i,j}^{+} = T^{[i+j]}_{|i-j| -1} (0) \comma \quad  c_{i,j}^{-} = T^{[i+j+1]}_{|i-j| -1} (0)
   \period
\eqe
Consequently, $\Delta A_{\rm BDS}$ is represented by 
the T-functions as
\eqb
    \Delta A_{\rm BDS} =  \frac{1}{4}\sum_{i,j=1}^{n} 
    \log \frac{T_{|i-j|-1}^{[i+j]}}{T_{|i-j-1|-1}^{[i+j+1]}}
     \log \frac{T_{|i-j-1|-1}^{[i+j]}}{T_{|i-j |-1}^{[i+j+1]}} \comma
\eqe
where  $T_{s}^{[k]}(\theta)$ are evaluated at $\theta =0$.
We see that each term in the remainder function (\ref{R2n})
nicely fits into the language of two-dimensional integrable models:
$A_{\rm free}$ is the free energy, and $A_{\rm periods}$ and $\Delta A_{\rm BDS}$
are given by the mass parameters  and the T-functions, respectively. 
Now, we have only to expand $T_{s}$ around the CFT limit. 

\subsection{Expansion of T-functions}

To expand the T-functions, we use an interesting 
relation\cite{Bazhanov:1994ft,Dorey:1999cj,Dorey:2005ak} between
the T-function and the boundary entropy or the $g$-function \cite{Affleck:1991tk}.
Since the $g$-function is regarded as a  boundary contribution 
to the free energy, one can compute it around the CFT limit by using CPT 
with boundary \cite{Dorey:1999cj,Dorey:2005ak}. 

The precise relation between the $g$- and T-functions is obtained by (i) constructing
the reflection factors of the HSG model which satisfy the unitarity,
the crossing symmetry, and the  boundary Yang-Baxter equation, (ii)
deriving the integral equations for the $g$-functions associated 
with the boundaries corresponding to the reflection factors, and (iii) comparing 
those equations with those for the T-functions. 
Here, to satisfy the boundary Yang-Baxter equation,   the imaginary 
resonance parameters or the phases of the mass parameters need to be vanishing, 
as they are in our discussion so far.
For details, we refer to Ref. \cite{Hatsuda:2011ke}. 

In expanding $T_{s}$, we then use their quasi-periodicity for $n$ odd, 
\eqb\label{Tperiod}
  T^{[n]}_{s}(\theta)= T_{n-2-s}(\theta) \period
\eqe
This  follows from the relations (\ref{YT}) which are written in the form of the T-system,
\eqb
  1+ T_{s+1}(\theta) T_{s-1}(\theta) = T_{s}^{[+1]}(\theta) T_{s}^{[-1]}(\theta) \comma
\eqe
with $T_{0} = T_{n-2} =1$. Combining this quasi-periodicity and  
the structure of the CPT,  the T-functions are expanded for real $m_{s}$ 
and $n$ odd as
\eqb\label{Texpand}
   T_{s}(\theta) = \sum_{p,q=0}  
   t^{(p,2q)}_{s} l^{(1-\Delta)(p+q)} \cosh \Bigl( \frac{2p}{n} \theta \Bigr) \comma
\eqe
with $t_{n-2-s}^{(p,2q)} = (-1)^{p} t_{s}^{(p,2q)}$. 
The T-system then fixes the lower coefficients as 
\eqb\label{ts00}
   t_{s}^{(0,0)} = \sin\Bigl( \frac{s+1}{n} \pi\Bigr) \Big/ \sin \Bigl( \frac{\pi}{n} \Bigr) \comma
\eqe
and $t_{s}^{(1,0)} = t_{s}^{(0,2)} = t_{s}^{(1,2)} = 0$.  Further 
using the results on the conformal perturbation of the $g$-function, 
and translating them into the expansion of  the T-functions,  we obtain 
\eqb\label{ts20}
   \frac{t_s^{(2,0)}}{t_{s}^{(0,0)}}=-\frac{ \kappa_{n}G 
   \cdot B(1-2\Delta,\Delta)}{2(2\pi)^{1-2\Delta}}
\left( \frac{\sin(\frac{3(s+1)\pi}{n})}{\sin(\frac{(s+1)\pi}{n})} 
\sqrt{\frac{\sin(\frac{\pi}{n})}{\sin(\frac{3\pi}{n})}}-
\sqrt{\frac{\sin(\frac{3\pi}{n})}{\sin(\frac{\pi}{n})}}\right) \comma
\eqe
where $B(a,b) =\Gamma(a) \Gamma(b)/\Gamma(a+b)$.
We note that the ratios of the sine functions come from the modular S-matrix 
of the coset CFT.
In relation to the $g$-function, the form of  $t_{s}^{(0,0)}$ in (\ref{ts00}) 
is also interpreted in this way.
From  the T-system, one can also show that $t_{s}^{(0,4)}$ are given by
$t_{s}^{(2,0)}$ via
\eqb\label{ts04}
  2t_s^{(0,0)}t_s^{(0,4)}+\frac{1}{2}(t_s^{(2,0)})^2\cos\Bigl( \frac{4\pi}{n} \Bigr)
=t_{s-1}^{(0,0)}t_{s+1}^{(0,4)}+t_{s+1}^{(0,0)}t_{s-1}^{(0,4)}
+\frac{1}{2}t_{s-1}^{(2,0)}t_{s+1}^{(2,0)} \period
\eqe
Together with (\ref{ts20}), we see that all $t_{s}^{(2,0)}$ and $t^{(0,4)}_{s}$
are expressed, e.g.,  by $t_{1}^{(2,0)} \propto \kappa_{n} G(\tilde{M}_{s})$. 

\subsection{Expansion of remainder function}

From the expansion of $A_{\rm free}$ and the one of $\Delta A_{\rm BDS}$ 
via those of 
$T_{s}$, the remainder function is expanded around the CFT limit. After some algebras,
we find that 
\eqb\label{R2nExpand}
  R_{2n} = R_{2n}^{(0)}+l^{\frac{8}{n}}R_{2n}^{(4)}+{\cal O}(l^{\frac{12}{n}}) \comma
\eqe
for $n$ odd  and real $m_{s}$, where 
\eqb\label{R2n04}
 R_{2n}^{(0)} & = & \frac{\pi}{4n}(n-2)(3n-2)-\frac{n}{2}\sum_{s=1}^{{(n-3)}/{2}} 
\log^2 \biggl( \frac{\sin (\frac{(s+1)\pi}{n})}{\sin (\frac{s\pi}{n})} \biggr) \comma  \nn \\
R_{2n}^{(4)} &=&\frac{\pi}{6}C_n^{(2)} {  \kappa_n^2 G^2(\tilde{M}_j) }
-\frac{n}{4}\Biggl[
\sum_{s=1}^{(n-3)/2} A_{n,s}-2 \biggl( \frac{t_{(n-3)/2}^{(2,0)}}{t_{(n-3)/2}^{(0,0)}}\biggr)^2
\sin^2\Bigl(\frac{\pi}{n}\Bigr)\Biggr] \comma
\eqe
and 
\eqb\label{Ans}
  A_{n,s} 
  &=& \left[ \biggl(\frac{t_{s-1}^{(2,0)}}{t_{s-1}^{(0,0)}}\biggr)^2
  +\biggl( \frac{t_{s}^{(2,0)}}{t_{s}^{(0,0)}}\biggr)^2 \right]
\cos\biggl(\frac{2\pi}{n} \biggr)
-\frac{2t_{s-1}^{(2,0)}t_s^{(2,0)}}{t_{s-1}^{(0,0)}t_{s}^{(0,0)}}\nonumber  \\
&& \ + \, \left[ \biggl( \frac{t_{s-1}^{(2,0)}}{t_{s-1}^{(0,0)}}\biggr)^2
-\biggl( \frac{t_{s}^{(2,0)}}{t_{s}^{(0,0)}}\biggr)^2-
4\biggl(\frac{t_{s-1}^{(0,4)}}{t_{s-1}^{(0,0)}}
-\frac{t_{s}^{(0,4)}}{t_{s}^{(0,0)}}\biggr)\right]
\log \biggl( \frac{t_s^{(0,0)}}{t_{s-1}^{(0,0)}} \biggr) 
 \period
\eqe
The first term in the expansion $R_{2n}^{(0)}$ gives the remainder function 
in the CFT limit, or for the regular-polygonal Wilson loops. We also note that  
$t_{s}^{(3,0)}, t_{s}^{(2,2)}$ and $t_{s}^{(4,0)}$ do not appear, and hence 
$R_{2n}^{(4)}$ is expressed by $t_{1}^{(2,0)}$ or $\kappa_{n} G$.

\subsection{$\bbZ_{2n}$-symmetry and remainder function for complex $m_{s}$}

The absence of $t_{s}^{(3,0)}, t_{s}^{(2,2)}, t_{s}^{(4,0)}$  in the above expansion 
is understood as a consequence of the $\bbZ_{2n}$-symmetry, which is the symmetry 
under the cyclic shift of  the cusp points: $k$-th  cusp $\to $ $(k+1)$-th cusp  
or, in terms of the light-cone coordinates, 
\eqb
  x_{j}^{-} \to x_{j+1}^{+} \comma \quad x_{j}^{+} \to x_{j}^{-} \period
\eqe
This is concisely expressed in terms of the Y-functions as \cite{Gaiotto:2010fk}
\eqb\label{Z2nY}
   Y_{s}(\theta) \to Y_{s}^{[+1]}(\theta) \comma
\eqe
or in terms of the mass parameters as $m_{s} \to m_{s}/i$.
Acting with this symmetry twice induces a simple translation 
$x_{j}^{\pm} \to x_{j+1}^{\pm}$.
For general complex $m_{s}$, the expansion of $T_{s}$ in (\ref{Texpand}) is
modified so that 
$t_{s}^{(p,2q)} \cosh (2p \theta/n)\to \frac{1}{2} ( t_{s}^{(p,2q)}  e^{2p \theta/n} 
+ \bar{t}_{s}^{(p,2q)}  e^{-2p \theta/n}) $. The shift $\theta \to \theta + \pi i/2$
 in (\ref{Z2nY})
is thus translated into the phase shift of the coefficients 
$t_{s}^{(p,2q)}, \bar{t}_{s}^{(p,2q)}$.
Then, the non-constant invariant combinations under the $\bbZ_{2n}$-symmetry 
are only $t_{s}^{(2,0)} \bar{t}_{s}^{(2,0)}$
and $t_{s}^{(0,4)}$ up to $\calO(l^{8 \over n})$, which explains the terms
in the expansion of $R_{2n}$.

So far, we have considered real $m_{s}$ corresponding to the vanishing
imaginary resonance parameters. The expansion for complex $m_{s}$
is also obtained so as to maintain this $\bbZ_{2n}$-symmetry:  we rewrite 
the expansion in terms of the 
$\bbZ_{2n}$-invariant  combinations of  $t_{s}^{(p,2q)}$, and then
continue the real $\tilde{M}_{s}$ in $t_{s}^{(p,2q)}$
to complex $\tilde{M}_{s}$.
Up to $\calO(l^{8 \over n})$, this procedure reduces to replacing 
$\kappa_{n}^{2} G^{2}$ for real $m_{s}$ by $\kappa_{n}^{2}G\bar{G}$
for complex $m_{s}$.
The $\bbZ_{2n}$-symmetry strongly constrains the structure of the remainder 
function in this way.

\subsection{Case of $n$ even}

So far, we have focused on the case of $n$ of $2n$ odd. For $n$ even, there are several
changes. For example, the remainder function has an extra term $A_{\rm extra}$ 
in addition to those in (\ref{R2n}) due to 
a non-trivial monodromy of an auxiliary variable defined by $ dw = \sqrt{p(z)} dz$.
The expressions of $A_{\rm periods}$ and $ \Delta A_{\rm BDS}$ are also changed.
Since $T_{n-2} \neq 1$ generally for $n$ even, the quasi-periodicity of the T-functions
is also modified. Due to this, their expansions have  extra factors compared with  
(\ref{Texpand}).  The remaining part, which we denote by $\hat{T}_{s} $,  however,
has the same quasi-periodicity and the form of the expansion
as in (\ref{Tperiod}) and  (\ref{Texpand}), respectively.
We refer to Ref. \cite{Hatsuda:2011jn} for derails. In any case,
since the extra term $A_{\rm extra}$  and the extra factors in $T_{s}$ are irrelevant 
up to $o(l)$,  the expansion of $R_{2n}$  up to $\calO(l^{\frac{8}{n}})$ for $n \geq 10$ 
is similar to the one for $n$ odd. We then  
obtain the expansion of the form  (\ref{R2nExpand})  with 
\eqb
    R_{2n}^{(0)}&=&\frac{\pi}{4n}(n-2)(3n-2)-\frac{n}{2}\sum_{s=1}^{n/2-1} 
\log^2 \biggl( \frac{\sin (\frac{(s+1)\pi}{n})}{\sin (\frac{s\pi}{n})} \biggr) \comma
 \nn \\
R_{2n}^{(4)}&=& \frac{\pi}{6}C_n^{(2)}\kappa_n^2 G^2(\tilde{M}_j)
-\frac{n}{4}
\sum_{s=1}^{n/2-1} \hat{A}_{n,s} \comma
\eqe
for $n \geq 10$.
Here, $\hat{A}_{n,s}$ are give by (\ref{Ans}) with $t^{(p,2q)}_{s}$ replaced by 
 $\hat{t}^{(p,2q)}_{s}$, which are the expansion coefficients of $\hat{T}_{s}$.
In the expansion up to this order, the relevant $\hat{t}^{(p,2q)}_{s}$ 
are however the same as 
$t^{(p,2q)}_{s}$ and given by (\ref{ts00}), (\ref{ts20}) and (\ref{ts04}). 

For $n=6,8$, we have also checked that various modifications
in each term in $R_{2n}$ eventually cancel each other, and we are left with
the same expression as for $n \geq 10$. The relevant coefficients 
$\hat{t}_{s}^{(p,2q)}$ are the same as  $t_{s}^{(p,2q)}$ and given by 
(\ref{ts00}), (\ref{ts20}) and (\ref{ts04}) again.
The case of $n=4$ is special in that the TBA system becomes trivial. 
In this case, the integral 
expression of the remainder function is given in Ref. \cite{Alday:2009yn}, 
and is expanded in $l^{2}$ to all orders in Ref. \cite{Hatsuda:2011ke}.

The expansion for general complex $m_{s}$ is obtained by complexifying $m_{s}$
similarly to the case of $n$ odd.

\subsection{Momentum dependence}

The expansion so far is given in terms of   
$ | t_{s}^{(2,0)}|^{2} \propto \kappa_{n}^{2} G \bar{G} $.
To express the remainder function as a function of  the momenta, we need to 
find the relation between those expansion parameters and the cross-ratios.
For this purpose, we first note that  the Y-functions have the quasi-periodicity, 
\eqb
    Y^{[n]}_{s}(\theta)= Y_{n-2-s}(\theta) \comma
\eqe
for $n$ both odd and even. This follows 
from the Y-system (\ref{Ysystem}). Then, similarly to $T_{s}$, 
the Y-functions for general complex $m_{s}$ are expanded  around the CFT limit as 
\eqb
   Y_{s}(\theta) = y_{s}^{(0,0)} + \frac{1}{2} \Bigl( y_{s}^{(2,0)} e^{{4 \over n} \theta} 
   + \bar{y}_{s}^{(2,0)} e^{- {4 \over n} \theta} \Bigr) l^{\frac{4}{n}}  
      + {\cal O}(l^{\frac{6}{n}}) \comma
\eqe
where  $y_{s}^{(0,0)}$ are the solution to the constant Y-system, 
\eqb
 y_{s}^{(0,0)} 
=  \sin \Bigl( \frac{s\pi}{n} \Bigr) \sin\Bigl( \frac{(s+2)\pi}{n} \Bigr) \Big/ 
\sin^{2} \Bigl( \frac{\pi}{n} \Bigr)
 \period
 \eqe
From the relations (\ref{YT}), we then find that
\eqb \label{eq:Ysk}
   Y_{s}^{[k]}(0) = y_{s}^{(0,0)}  + \cos\Bigl( \frac{2\pi}{n} \Bigr) t_{s}^{(0,0)}
    \Bigl( t_{s}^{(2,0)} e^{{2\pi \over n} ki } + 
      \bar{t}_{s}^{(2,0)} e^{- {2\pi \over n} ki} \Bigr) l^{\frac{4}{n}}  
      + {\cal O}(l^{\frac{6}{n}}) \period
\eqe
Inverting this gives
\eqb\label{tvarphi}
   |t_{s}^{(2,0)}|   l^{\frac{4}{n}} &=& { \delta Y_{s}^{[0]}  \over 2 t_{s}^{(0,0)}
  \cos \bigl(\frac{2\pi}{n} \bigr) \cos \phi_{s} }
   \comma \nn \\
   {2\pi \over n} \phi_{s} &=& \arctan\Bigl( \cot\Bigl(\frac{2\pi}{n} \Bigr)
   \frac{\delta Y_{s}^{[-1]} - \delta Y_{s}^{[1]}}{\delta Y_{s}^{[-1]} 
   + \delta Y_{s}^{[1]}} \Bigr) 
   \comma 
\eqe
up to the present order, where we have set $ t_{s}^{(2,0)} 
= |t_{s}^{(2,0)}| e^{i\phi_{s}} $, and  
$\delta Y_{s}^{[k]}$ are the deviations of the cross-ratios
from the CFT/regular-polygonal limit,  
$ \delta Y_{s}^{[k]} := Y_{s}^{[k]} - y_{s}^{(0,0)}$.
By using the relations (\ref{Ycrossratio}), 
these are  indeed expressed in terms of the cross-ratios
(which depend on each other at this order through (\ref{eq:Ysk})).
The momentum dependence of the remainder function is found by 
substituting (\ref{tvarphi}) into the expansion.

\subsection{Mass-coupling relations}

At higher orders,
the expansion may not be expressed only 
by $\kappa_{n} G$. Thus, we need to find the precise form of the perturbing 
term  $ \kappa_{n} \Phi$ for given $m_{s}$, i.e., the mass-coupling relation 
in the HSG model. This is also necessary to make contact with numerics  
using the TBA system (\ref{AdS3TBA}) for given masses, as well as to 
find the connection between $m_{s}$ and the cross-ratios/shape 
of the minimal surfaces.

This is achieved for some cases 
where the TBA system has only one mass scale. A classification 
of such cases is given in Ref. \cite{Ravanini:1992fi}.
In this subsection, we set $m_{s}$ to be real again. 
To see this, we first parameterize $\Phi$ as 
\eqb
    \Phi = \sum_{l,\bar{l} = 1}^{n-3} (\bs{\lambda})^l 
    ({\bs{\lambda}})^{\bar{l}} \phi_{l,\bar{l}}
    \comma
\eqe
where $\phi_{l,l'}$ are the weight 0 adjoint operators normalized so that 
$ \langle  \phi_{l,\bar{l}} (z) \phi_{l',\bar{l}'} (0)  \ket   
=  \delta_{l,l'} \delta_{\bar{l},\bar{l}'}|z|^{-4\Delta} $, and 
$\bs{\lambda} =
\sum_{j=1}^{n-3} \tilde{M}_j^{1-\Delta} \bs{\hlambda}_j$.
Classically, $\hat{\bs{\lambda}}_{j}$ are the fundamental weights 
of $su(n-2)$ \cite{FernandezPousa:1997iu,Dorey:2004qc}.
In this parameterization,  one has 
\eqb
   G(\tilM_{j}) = \sum_{i,j=1}^{n-1}
   \tilM_{i}^{\frac{2}{n}}F_{ij}  \tilM_{j}^{\frac{2}{n}} \comma \quad 
   F_{ij} := \hat{\bs{\lambda}}_{i} \cdot \hat{\bs{\lambda}}_{j} \period
\eqe

When the mass parameters are set to be
$M_{1} = M,   \, M_{2} = \cdots = M_{n-3} = 0$,
the TBA system reduces to the one for the (RSOS)$_{n-2}$ 
scattering theory \cite{Zamolodchikov:1991vh,Itoyama:1990pv},  which 
is regarded as a massive integrable deformation of the unitary minimal model 
${\cal M}_{n-1,n}$ by the $\phi_{1,3}$ operator. 
From the result in Ref. \cite{Zamolodchikov:1995xk}, we then find that
\eqb\label{Mn-1}
    \kappa_{n} F_{11} 
   &=& \frac{1}{\pi} \frac{n^{2}}{(n-2)(2n-3)} 
   \biggl[ \gamma\Bigl( \frac{3(n-1)}{n}\Bigr) \gamma\Bigl( \frac{n-1}{n}\Bigr)
   \biggr]^{\frac{1}{2}}
   \biggl[ \frac{\sqrt{\pi}\Gamma(\frac{n}{2})}{2\Gamma(\frac{n-1}{2})}
   \biggr]^{ \frac{4}{n} } \period
\eqe
When only the $k$-th mass parameter is non-vanishing, i.e., 
$M_{j} = \delta_{jk} M$,
the TBA system reduces to the one for an integrable deformation  
of the coset CFT associated
with $\widehat{su}(2)_{k}\oplus \widehat{su}(2)_{n-2-k}/\widehat{su}(2)_{n-2}$  by the 
$\phi_{1,1,{\rm adj}}$ operator \cite{Zamolodchikov:1991vg}.
Using the result  in Ref. \cite{Fateev:1993av}
together with (\ref{Mn-1}), we also find that
\eqb\label{su2coset}
 \frac{F_{kk}}{F_{11}}=\frac{k(n-k-2)}{n-3}
\left[ \frac{\sqrt{\pi}}{2} \frac{\Gamma(\frac{n-1}{2})}{\Gamma(\frac{k}{2}+1)
\Gamma(\frac{n-k}{2})}
\right]^{\frac{4}{n}} \period
\eqe
Furthermore, when $M_1=M_{n-3}=M, \,  M_2=\dots=M_{n-4}=0$ for $n$ odd,
the TBA system reduces to the one for the magnonic 
$T_{(n-3)/2} = A_{n-3}/\bbZ_{2}$ system, which is described by 
an integrable perturbation of the non-unitary coset CFT associated with 
$\widehat{su}(2)_{n/2-3} \oplus \widehat{su}(2)_{1}/\widehat{su}(2)_{n/2-2}$  
by the $\phi_{1,1,{\rm adj}}$ operator, or
the non-unitary minimal model ${\cal M}_{n-2,n}$ perturbed by the $\phi_{1,3}$ 
operator \cite{Ravanini:1992fi}.
Then, using the result again in Ref. \cite{Fateev:1993av}, one can derive 
\eqb\label{Mn-2n}
  1+\frac{F_{1,n-3}}{F_{11}}=\frac{n-2}{n-3} \left[
\frac{\Gamma(\frac{n}{4})\Gamma(\frac{n-1}{2})}
{\Gamma(\frac{n}{4}-\frac{1}{2})\Gamma(\frac{n}{2})} \right]^{\frac{4}{n}}  \period
\eqe

From (\ref{Mn-1}), (\ref{su2coset}) and (\ref{Mn-2n}), the form of 
$\kappa_{n} \Phi$  is fixed
in the corresponding cases. By changing the mass scale $M$, one can  trace the
the remainder function along the trajectories in the momentum space parameterized
by the mass parameters. As we will discuss shortly,
these data completely fix the mass-coupling relation for the 10-point amplitudes.

\section{Lower-point cases}

So far, we have discussed the expansion of the remainder function for 
general $n$ of $2n$. In this section,
we specialize to the case of $2n=8$ and $10$ \cite{Hatsuda:2011ke}.

\subsection{Eight-point amplitudes}

In the case of the eight-point amplitudes, 
the HSG model reduces to the Ising model and the TBA system becomes trivial. 
With the help of the results on the free energy and the $g$-function 
for the Ising model \cite{Klassen:1990dx,Dorey:2004xk},
one can derive 
the all-order expansion of the integral expression \cite{Alday:2009yn} of
the remainder function:  
\eqb
 R_8&=& 
\frac{5\pi}{4}-\frac{1}{2}\log\(2\cosh \frac{l\cos\varphi}{2}\)
\log\(2\cosh \frac{l\sin\varphi}{2}\)+\frac{l^2}{8\pi} \nn  \\
&+ &  \pi \sum_{k=1}^\infty 
 { \frac{1}{2} \choose k+1 }
 \(1-\frac{1}{2^{2k+1}}\)\zeta(2k+1)\(1-\frac{k+1}{2k+1}
f_k(\varphi)\)\(\frac{l}{\pi}\)^{2k+2}
\comma \quad 
\eqe
where $\zeta(z)$ is the zeta-function,  $l, \varphi$ are given by 
the mass parameter as 
\eqb
     m = l e^{i\varphi} \comma
\eqe
and the function $f_{k}(\varphi)$ are given through the hypergeometric function as 
\eqb
 f_k(\varphi) := 
 \cos^2\varphi\; {}_2F_1(-k,1;\frac{1}{2}-k;\sin^2\varphi)
 +\sin^2\varphi\; {}_2F_1(-k,1;\frac{1}{2}-k;\cos^2\varphi) \period
\eqe
We have refrained from expanding the logarithmic terms. 

\subsection{Ten-point amplitudes}

For the ten-point amplitudes, there are two mass parameters $m_{1}, m_{2}$.
In this case, the consideration in the previous section completely determines 
the expansion up to and including terms of  ${\cal O}(l^{4(1-\Delta)})$:
\eqb\label{R10}
   R_{10} 
     =  R_{10}^{(0)} + R_{10}^{(4)} \cdot l^{8/5} + {\cal O}(l^{12/5}) \comma
\eqe
where
\eqb
     R_{10}^{(0)} &= &\frac{39}{20}\pi-\frac{5}{2}\log^2\bigl( 2\cos \frac{\pi}{5} \bigr) 
     \comma  \nonumber \\
   R_{10}^{(4)} &=&  \bigl( -\frac{1}{5}\tan \frac{\pi}{5} +C_1 \bigr)  \cdot \bigl|  
    t_{1}^{(2,0)}  \bigr|^{2} \comma
\eqe
with 
\eqb\label{t1Cs}
    t_{1}^{(2,0)} &=&  C_{2} (\tilde{M}_1^{4/5}+\tilde{M}_2^{4/5}-C_{3} 
    \tilde{M}_1^{2/5}\tilde{M}_2^{2/5})  \comma \nn \\
   &&  C_{1}= 20\cos^4\bigl(\frac{2\pi}{5} \bigr)
 \Bigl( 1- 5^{-1/2} \log\bigl( 2\cos \frac{\pi}{5} \bigr) \Bigr) \comma
 \nonumber \\
&& 
 C_{2} = \frac{1}{4 \cdot 6^{1/5}}\Gamma\bigl(-1/5 \bigr)
\Bigl[ 10\cos \frac{\pi}{5} \gamma\bigl(3/5 \bigr)\gamma\bigl(4/5 \bigr) \Bigr]^{1/2} 
\comma \nonumber \\
&& 
 C_{3} =2-\bigl(\frac{3}{\pi^2}\bigr)^{1/5}\gamma\bigl(1/4\bigr)^{4/5} \comma 
 \eqe
and $\gamma(x) = \Gamma(x)/\Gamma(1-x)$. This holds for the complex 
mass parameters,
\eqb
   m_{s} = \tilde{M}_{s} l =  e^{i\varphi_{s}} | \tilde{M}_{s} |  \, l \comma
\eqe
and $t_{1}^{(2,0)}$ is related to $\kappa_{n} G(\tilde{M}_{s})$ 
by (\ref{ts20}). 
 
The results can be compared with the numerical results 
from the TBA equations (\ref{AdS3TBA}).
We show  comparisons of the remainder function
 for various complex $m_{s}$ in Fig.~6. We find a good agreement
between our analytic expansions and  the numerics.
\begin{figure}[t]
   \bc
   \includegraphics*[width=6.5cm]{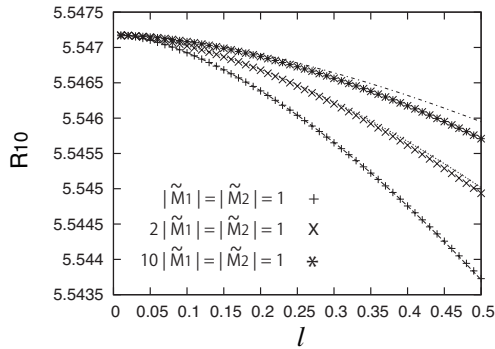}
 \vspace*{2pt}
 \caption{Comparisons of analytic results at strong coupling 
 and numerics for various complex 
 mass parameters with $\varphi_{1} = \pi/20, \varphi_{2} = \pi/5$.
 Dashed lines represent the analytic expansion (\ref{R10})-(\ref{t1Cs}),
 whereas the points ($+, \times, \ast$) are from numerics.
 }
 \ec
\end{figure}
\section{Comparison with two-loop results}

For the amplitudes corresponding to the minimal surfaces in $AdS_{3}$,
the analytic expression  of the remainder function has been given  
at two loops \cite{Heslop:2010kq}.
In this section, we compare this two-loop remainder function with that at strong 
coupling for  the same cross-ratios \cite{Hatsuda:2011ke,Hatsuda:2011jn}. 

First, by expressing the cross-ratios
by the Y-/T-functions, and then substituting their expansions into  the two-loop result, 
one finds an expansion similar to (\ref{R2nExpand}): 
\eqb
   R^{\rm 2\mbox{-}loop}_{2n}   = R_{2n}^{\rm 2\mbox{-}loop \, (0)} 
   + l^{\frac{8}{n}} R_{2n}^{\rm 2\mbox{-}loop \, (4)} 
+ {\cal O}(l^{\frac{12}{n}}) \period
\eqe
The overall coupling constant $\lambda^{2}$ has been omitted above.
For comparison, we next introduce a shifted and rescaled remainder function 
following Ref. \cite{Brandhuber:2009da}, 
\eqb
  \bar{R}_{2n}^{\rm 2\mbox{-}loop} := 
  \frac{R_{2n}^{\rm 2\mbox{-}loop}-R_{2n,{\rm UV}}^{\rm 2\mbox{-}loop}}{
R_{2n, {\rm UV}}^{\rm 2\mbox{-}loop}-(n-2)R_{6}^{\rm 2\mbox{-}loop} } \comma
\eqe
where $R_{2n, {\rm UV}}^{\rm 2\mbox{-}loop}$ are the remainder functions 
in the CFT/UV limit.  This rescaled remainder function is calibrated so that
it approaches 0 in the UV limit ($l \to 0$) and $-1$ in the opposite IR limit ($l \to \infty$).
The rescaled remainder function at strong coupling, $\bar{R}^{\rm strong}_{2n}$,  
is defined similarly.
It has been observed numerically 
that those remainder functions are
close to each other for $2n=8$  \cite{Brandhuber:2009da} .

Since the dependence on the mass parameters/momenta in the expansion
up to ${\cal O}(l^{4(1-\Delta)})$ is encoded in only one function, e.g.,  
$t_{1}^{(2,0)} \propto \kappa_{n} G$,
the ratio of the rescaled remainder functions at strong coupling and at two loops becomes 
just a number up to this order. For example, for lower-point amplitudes, we find that
\eqb
 \bar{\rho}_{8}  \ \approx  \ 1.0257 
  \comma  &\qquad & \bar{\rho}_{10}  \  \approx  \ 0.9841 \comma
  \nonumber \\
   \bar{\rho}_{12} \  \approx  \  0.9609 \comma
    &&
   \bar{\rho}_{14}  \  \approx  \  0.9463 \comma
    \nonumber \\
   \bar{\rho}_{16} \  \approx  \  0.9366 \comma
   && \bar{\rho}_{18}  \  \approx  \  0.9297 \comma
\eqe
where 
$\bar{\rho}_{2n} := 
{\bar{R}^{\rm strong}_{2n}}/{\bar{R}^{\rm 2\mbox{-}loop}_{2n}} $.
After some analysis, one can also find the asymptotic behavior,  
\eqb
     \bar{\rho}_{2n} \to 0.905 - \frac{0.118}{n} \quad (n \gg 1) \period
\eqe 
In principle, the ratio may take  any values. 
We would thus like to say that the remainder functions at strong coupling and 
at two loops are (surprisingly) close for all $2n$, though different. 
\section{Summary}

In this talk, we discussed  gluon scattering amplitudes/null-polygonal Wilson
loops of four-dimensional ${\cal N} = 4$ SYM at strong coupling. 
By the AdS/CFT correspondence,
they are given by the minimal surfaces in $AdS_{5}$. Those minimal surfaces
are in turn described by a set of integral equations. When the surfaces are contained 
in $AdS_{3}$ or $AdS_{4}$ subspace, those integral equations are identified
with the TBA equations of the HSG model, which is an integrable deformation 
of the gauged WZW model by the weight 0 adjoint operators. 

Based on the relation between the amplitudes and this underlying 
two-dimensional integrable model, 
we derived analytic expansions of the amplitudes/Wilson loops corresponding 
to the minimal surfaces in $AdS_{3}$ around the CFT/regular-polygonal limit.
There were two non-trivial terms to be expanded in the remainder function,
$A_{\rm free}$ and $\Delta A_{\rm BDS}$.
The former, $A_{\rm free}$,  is nothing but the free energy of the 
HSG model, and is expanded  around the CFT limit by conformal perturbation.
For the latter, $\Delta A_{\rm  BDS}$, 
the sequential cross-ratios $c_{i,j}^{\pm}$ therein 
are directly expressed by the T-functions. This implies  that every term 
in the remainder function
nicely fits into the language of the integrable model. Interestingly, these T-functions 
are related to the $g$-function/boundary entropy, and the $g$-function, which is regarded 
as a boundary contribution to the free energy, is expanded around the CFT limit 
by conformal perturbation with boundary.
Combining all the expansions together, we obtained the expansion 
of the remainder function.
We also compared our analytic expansions at strong coupling  with those at two loops.
We found that the rescaled remainder functions at strong coupling and at two loops 
are close to each other for all $2n$-point amplitudes.

For future, there still remain many questions to be addressed. For example, we still do not
find a physical reason why the minimal surfaces are described by 
the TBA equations. The situation would be similar to the case of the ODE/IM
correspondence \cite{Dorey:2007zx}.  
We do not find a physical reason for the relation between the $g$- and
T-functions, either. It is also not clear why the rescaled remainder functions are 
so close to each other at strong coupling and at two loops. 
This seems to be suggesting  some mechanism 
to constrain the amplitudes which is not yet understood. The understanding of 
such a mechanism, if any, would be important to uncover the full structure of 
the amplitudes to all orders. 

Our analysis may be extended to the case of the amplitudes corresponding to the
minimal surfaces in $AdS_{4}$. The integrable
model for the general case of $AdS_{5}$, however,
 has not been identified. The precise relation
between the perturbing operator and the mass parameters, i.e., the mass-coupling
relation, which connects the conformal perturbation, the TBA system 
and the minimal surfaces/amplitudes, has not yet been obtained generally.
These would be  important questions for future. 
In this talk, we discussed the amplitudes in the  strong-coupling limit. 
It would also be very interesting if one could incorporate the strong-coupling 
corrections and to interpolate the result to the weak-coupling side, as was the case
for the spectral problem.

I would like to end this talk by showing one picture (Fig.~7).
In this way, we see very interesting connections surrounding the gluon scattering
amplitudes among the ten-dimensional
string theory on $AdS_{5} \times S^{5}$, the four-dimensional SYM,
and the two-dimensional integrable systems (HSG model and Hitchin system) 
 and CFTs. It would be interesting to explore these connections further.
\begin{figure}[t]
   \bc
   \includegraphics*[width=11cm]{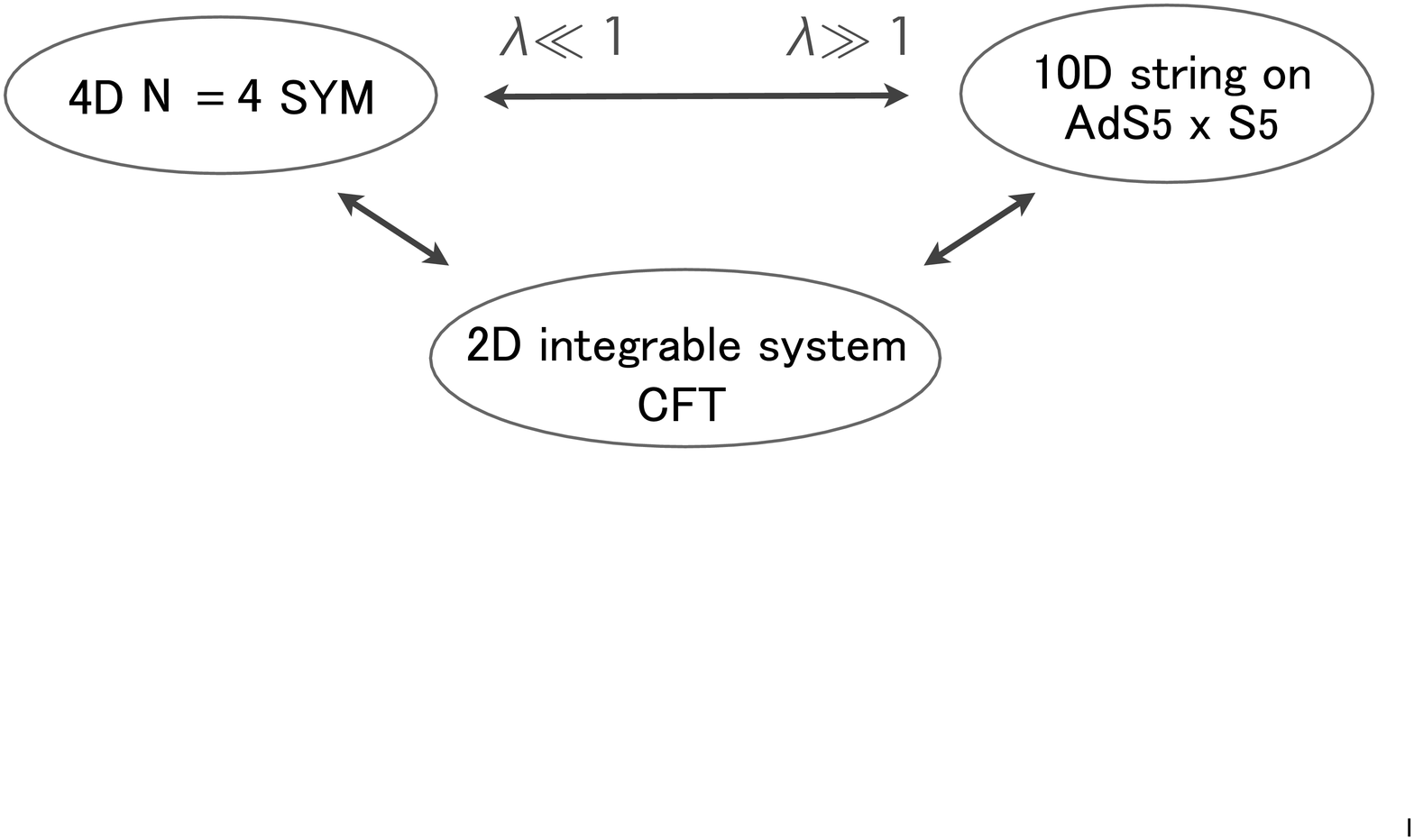}
 \vspace*{-70pt}
 \caption{Two-, four- and ten-dimensional theories surrounding 
 gluon scattering amplitudes.}
 \ec
\end{figure}

\section*{Acknowledgments}

The author would like to thank the organizers for the opportunity to speak
in this workshop. He would also like to thank Y. Hatsuda, K. Ito and K. Sakai
for fruitful collaboration. His work is supported in part by Grant-in-Aid for 
Scientific Research from  Ministry of Education, Culture, Sports, 
Science and Technology of Japan.

%
%
\def\thebibliography#1{\list
 {[\arabic{enumi}]}{\settowidth\labelwidth{[#1]}\leftmargin\labelwidth
  \advance\leftmargin\labelsep
  \usecounter{enumi}}
  \def\newblock{\hskip .11em plus .33em minus .07em}
  \sloppy\clubpenalty4000\widowpenalty4000
  \sfcode`\.=1000\relax}
 \let\endthebibliography=\endlist
%
%
\vspace{3ex}
\begin{center}
 {\large\bf References}
\end{center}
\par 

%

%

\begin{thebibliography}{999}
\parskip=-1.5pt

\bibitem{Hatsuda:2010cc}
 Y.~Hatsuda, K.~Ito, K.~Sakai and Y.~Satoh,
 JHEP {\bf 1004} (2010) 108.
 
\bibitem{Hatsuda:2010vr}
 Y.~Hatsuda, K.~Ito, K.~Sakai and Y.~Satoh,
 JHEP {\bf 1009 } (2010)  064. 
 
\bibitem{Hatsuda:2011ke}
 Y.~Hatsuda, K.~Ito, K.~Sakai and Y.~Satoh,
 JHEP {\bf 1104} (2011) 100.
  
\bibitem{Hatsuda:2011jn}
  Y.~Hatsuda, K.~Ito and Y.~Satoh,
  JHEP {\bf 1202} (2012) 003.

\bibitem{Minahan:2002ve}
  J.~A.~Minahan and K.~Zarembo,
  JHEP {\bf 0303} (2003) 013.

\bibitem{Bena:2003wd}
  I.~Bena, J.~Polchinski and R.~Roiban,
  Phys.\ Rev.\ D {\bf 69} (2004) 046002.

\bibitem{Beisert:2010jr}
  N.~Beisert, C.~Ahn, L.~F.~Alday, Z.~Bajnok, J.~M.~Drummond, L.~Freyhult, N.~Gromov 
  and R.~A.~Janik {\it et al.},
  Lett.\ Math.\ Phys.\  {\bf 99} (2012) 3.
          
\bibitem{Zamolodchikov:1989cf}
 Al.~B.~Zamolodchikov,
 Nucl.\ Phys.\  B {\bf 342 } (1990)  695.

\bibitem{Bajnok:2009vm}
  Z.~Bajnok, A.~Hegedus, R.~A.~Janik and T.~Lukowski,
  Nucl.\ Phys.\ B {\bf 827} (2010) 426.
  
\bibitem{Eden:2012fe}
  B.~Eden, P.~Heslop, G.~P.~Korchemsky, V.~A.~Smirnov and E.~Sokatchev,
  Nucl.\ Phys.\ B {\bf 862} (2012) 123.
   
\bibitem{Alday:2007hr}
L.~F.~Alday and J.~M.~Maldacena,
JHEP {\bf 0706} (2007)  064.
    
\bibitem{Alday:2009yn}
 L.~F.~Alday and J.~Maldacena,
 JHEP {\bf 0911} (2009) 082.
  
\bibitem{Alday:2009dv}
  L.~F.~Alday, D.~Gaiotto and J.~Maldacena,
  JHEP {\bf 1109} (2011) 032.

\bibitem{Alday:2010vh}
 L.~F.~Alday, J.~Maldacena, A.~Sever and P.~Vieira,
 J.\ Phys.\ A  {\bf 43} (2010) 485401.

\bibitem{Alday:2008yw}
  L.~F.~Alday and R.~Roiban,
  Phys.\ Rept.\  {\bf 468} (2008) 153.

\bibitem{Drummond:2007aua}
J.M.~Drummond, G.P.~Korchemsky and E. Sokatchev,
 Nucl.\ Phys.\  B {\bf 795} (2008) 385.
   
\bibitem{Bern:2005iz}
 Z.~Bern, L.~J.~Dixon and V.~A.~Smirnov,
 Phys.\ Rev.\  D {\bf 72} (2005) 085001.
 
\bibitem{Alday:2007he}
 L.~F.~Alday and J.~Maldacena,
 JHEP {\bf 0711} (2007) 068.
 
\bibitem{Drummond:2007au}
 J.~M.~Drummond, J.~Henn, G.~P.~Korchemsky and E.~Sokatchev,
 Nucl.\ Phys.\  B {\bf 826} (2010) 337.
 
\bibitem{Bern:2008ap}
 Z.~Bern, L.~J.~Dixon, D.~A.~Kosower, R.~Roiban, M.~Spradlin, C.~Vergu and A.~Volovich,
 Phys.\ Rev.\  D {\bf 78} (2008) 045007.
   
\bibitem{Drummond:2008aq}
 J.~M.~Drummond, J.~Henn, G.~P.~Korchemsky and E.~Sokatchev,
 Nucl.\ Phys.\  B {\bf 815} (2009) 142.

\bibitem{Sakai:2009ut}
  K.~Sakai and Y.~Satoh,
  JHEP {\bf 0910} (2009) 001.
 
\bibitem{Sakai:2010eh}
  K.~Sakai and Y.~Satoh,
  JHEP {\bf 1003} (2010) 077.

\bibitem{Gaiotto:2008cd}
  D.~Gaiotto, G.~W.~Moore and A.~Neitzke,
  Commun.\ Math.\ Phys.\  {\bf 299} (2010) 163.
  
\bibitem{Koberle:1979sg}
  R.~Koberle and J.~A.~Swieca,
  Phys.\ Lett.\ B {\bf 86} (1979) 209.
         
\bibitem{FernandezPousa:1996hi}
 C.~R.~Fernandez-Pousa, M.~V.~Gallas, T.~J.~Hollowood, J.~L.~Miramontes,
 Nucl.\ Phys.\  B {\bf 484 } (1997)  609. 

\bibitem{Zamolodchikov:1991et}
 Al.~B.~Zamolodchikov,
 Phys.\ Lett.\  B {\bf 253 } (1991)  391.
 
\bibitem{DelDuca:2010zg}
 V.~Del Duca, C.~Duhr and V.~A.~Smirnov,
 JHEP {\bf 1005} (2010) 084.
 
\bibitem{Goncharov:2010jf}
 A.~B.~Goncharov, M.~Spradlin, C.~Vergu and A.~Volovich,
 Phys.\ Rev.\ Lett.\  {\bf 105} (2010) 151605.
 
\bibitem{DelDuca:2010zp}
 V.~Del Duca, C.~Duhr and V.~A.~Smirnov,
 JHEP {\bf 1009} (2010) 015.
  
\bibitem{Heslop:2010kq}
 P.~Heslop and V.~V.~Khoze,
 JHEP {\bf 1011} (2010) 035.

\bibitem{Kuniba:2010ir}
 A.~Kuniba, T.~Nakanishi, J.~Suzuki,
 J.\ Phys.\ A {\bf 44 } (2011)  103001. 

\bibitem{Bazhanov:1994ft}
V.~V.~Bazhanov, S.~L.~Lukyanov and A.~B.~Zamolodchikov,
Commun.\ Math.\ Phys.\  {\bf 177} (1996) 381.

\bibitem{Dorey:1999cj}
P.~Dorey, I.~Runkel, R.~Tateo and G.~Watts,
Nucl.\ Phys.\  B {\bf 578} (2000) 85.

\bibitem{Dorey:2005ak}
P.~Dorey, A.~Lishman, C.~Rim and R.~Tateo,
Nucl.\ Phys.\  B {\bf 744} (2006) 239.

\bibitem{Affleck:1991tk}
 I.~Affleck, A.~W.~W.~Ludwig,
 Phys.\ Rev.\ Lett.\  {\bf 67 } (1991)  161.

\bibitem{Gaiotto:2010fk}
 D.~Gaiotto, J.~Maldacena, A.~Sever and P.~Vieira,
 JHEP {\bf 1103} (2011) 092.

\bibitem{Ravanini:1992fi}
 F.~Ravanini, R.~Tateo, A.~Valleriani,
 Int.\ J.\ Mod.\ Phys.\  A {\bf 8 } (1993)  1707. 

\bibitem{FernandezPousa:1997iu}
  C.~R.~Fernandez-Pousa and J.~L.~Miramontes,
  Nucl.\ Phys.\ B {\bf 518} (1998) 745.
 
\bibitem{Dorey:2004qc}
P.~Dorey and J.~L.~Miramontes,
Nucl.\ Phys.\  B {\bf 697} (2004) 405.

\bibitem{Zamolodchikov:1991vh}
 Al.~B.~Zamolodchikov,
 Nucl.\ Phys.\  B {\bf 358 } (1991)  497.

\bibitem{Itoyama:1990pv}
 H.~Itoyama, P.~Moxhay,
 Phys.\ Rev.\ Lett.\  {\bf 65 } (1990)  2102.

\bibitem{Zamolodchikov:1995xk}
Al.~B.~Zamolodchikov,
Int.\ J.\ Mod.\ Phys.\  A {\bf 10} (1995) 1125.

\bibitem{Zamolodchikov:1991vg}
  Al.~B.~Zamolodchikov,
 Nucl.\ Phys.\  B {\bf 366}  (1991) 122.
 
\bibitem{Fateev:1993av}
V.~A.~Fateev,
Phys.\ Lett.\  B {\bf 324} (1994) 45.
 
\bibitem{Klassen:1990dx}
 T.~R.~Klassen and E.~Melzer,
 Nucl.\ Phys.\  B {\bf 350} (1991) 635.

\bibitem{Dorey:2004xk}
 P.~Dorey, D.~Fioravanti, C.~Rim and R.~Tateo,
 Nucl.\ Phys.\  B {\bf 696} (2004) 445.
  
\bibitem{Brandhuber:2009da}
A.~Brandhuber, P.~Heslop, V.~V.~Khoze and G.~Travaglini,
JHEP {\bf 1001} (2010) 050.

\bibitem{Dorey:2007zx}
  P.~Dorey, C.~Dunning and R.~Tateo,
  J.\ Phys.\ A  {\bf 40} (2007) R205.

\end{thebibliography}
\end{document}